\documentclass[12pt]{article}
\usepackage[margin=1 in]{geometry} 
\usepackage{amsmath}
\usepackage{amsthm}
\usepackage{fancyhdr}
\usepackage{mathrsfs}
\usepackage{slashbox}
\usepackage{enumerate}
\usepackage{times}
\usepackage{epsfig}
\usepackage{amssymb}
\usepackage{setspace}
\usepackage{multirow}
\usepackage{graphicx}
\usepackage{latexsym}
\usepackage{float}
\usepackage[round]{natbib} 

\newtheorem{prop}{Proposition}
\newtheorem{corol}{Corollary}
\newtheorem{thrm}{Theorem}

\begin{document} 
\jot=10pt

\begin{center} 
\Large Marginally Interpretable Generalized Linear Mixed Models \\
\large Jeffrey J. Gory, Peter F. Craigmile, and Steven N. MacEachern \\
\normalsize Department of Statistics, The Ohio State University, Columbus, OH 43210
\end{center} 

\begin{abstract}
\singlespacing
Two popular approaches for relating correlated measurements of a non-Gaussian response variable to a set of predictors are to fit a \emph{marginal model} using generalized estimating equations and to fit a \emph{generalized linear mixed model} by introducing latent random variables.  The first approach is effective for parameter estimation, but leaves one without a formal model for the data with which to assess quality of fit or make predictions for future observations.  The second approach overcomes the deficiencies of the first, but leads to parameter estimates that must be interpreted conditional on the latent variables.  Further complicating matters, obtaining marginal summaries from a generalized linear mixed model often requires evaluation of an analytically intractable integral or use of attenuation factors that are not exact.  We define a class of marginally interpretable generalized linear mixed models that lead to parameter estimates with a marginal interpretation while maintaining the desirable statistical properties of a conditionally-specified model.  We discuss the form of these models under various common link functions and also address computational issues associated with these models.  For logistic mixed effects models, we introduce an accurate and efficient method for evaluating the logistic-normal integral.
\end{abstract}

\vspace{-0.2in}
\section{Introduction} \label{sec:introduction}
\vskip-0.05in
\hspace{.9cm} The generalized linear mixed model (GLMM) combines the generalized linear model \citep[see][]{nelderwedderburn1972, mccullaghnelder1989} with the linear mixed model \citep[see][]{hendersonetal1959, henderson1975, lairdware1982} to form a model that allows for non-Gaussian responses as well as random effects.  Such a model relates a linear combination of $p$ predictors $\mathbf{x}$ and $q$ random effects $\mathbf{U}$ to a response $Y$ through a \emph{link function} $g(\cdot)$ \citep[see][]{stiratellietal1984, gilmouretal1985, mccullochetal2008}.  It has the form
\vspace{-0.2 in}
\begin{equation*} 
\mu = \mathrm{E}(Y|\mathbf{U}=\mathbf{u}) = h(\mathbf{x}^T\boldsymbol{\beta}+\mathbf{d}^T\mathbf{u}),
\end{equation*}  
\vskip-0.2in \noindent
where $\boldsymbol{\beta}$ is a $p$-vector of fixed effect parameters, $\mathbf{d}$ is a $q$-vector of covariates, and $h(\cdot)=g^{-1}(\cdot)$ is the inverse link function.  To complete the specification of the model one typically assumes a distributional form for the random effects $\mathbf{U}$ and the response $Y$ given $\mathbf{U}$.  This defines a \emph{conditional model} for which the mean $\mu$ is conditioned on the random effects $\mathbf{U}$.

Interest often lies in \emph{marginal} or \emph{population-averaged} effects rather than conditional effects.  Although one could obtain marginal predictions from a GLMM, it is common to directly model the marginal mean using what is known as a \emph{marginal model}.  Such a model involves specification of a mean structure, typically written as $\mu=\mathrm{E}(Y)=h(\mathbf{x}^T\boldsymbol{\beta})$, and a covariance structure, usually with no distributional form explicitly assumed for the data.  Estimation of $\boldsymbol{\beta}$ for this model, ordinarily accomplished using \emph{generalized estimating equations} (GEE), can be robust to misspecification of the covariance structure \citep{liangzeger1986, zegerliang1986}.  However, a purely marginal model is typically not a fully-specified generative model, which makes it difficult to check and compare models and to make predictions at the individual level.

The distinction between marginal and conditional models is important because the parameters in these two types of models are generally not the same when the link function $g(\cdot)$ is nonlinear \citep[see][]{zegeretal1988, neuhausetal1991, neuhausjewell1993, diggleetal2002, ritzspiegelman2004}.  Several authors have addressed this discrepancy by attempting to find a proportional relationship between the marginal and conditional parameters.  For example, \citet{wanglouis2003} proposed a \emph{bridge distribution} $f_U$ for univariate random effects, which depends on the choice of inverse link function, such that for some constant $c$ and all $\mathbf{x}^T\boldsymbol{\beta} \in \mathbb{R}$ 
\vskip -0.2 in
\begin{equation} \label{proportionality}
\int h(\mathbf{x}^T\boldsymbol{\beta}+u)f_U(u) du = h(c \ \mathbf{x}^T\boldsymbol{\beta}).
\end{equation}
Similarly, \citet{caffoetal2007} proposed altering the inverse link function $h(\cdot)$ in order to satisfy (\ref{proportionality}).  Earlier work by \citet{zegeretal1988} and \citet{neuhausetal1991} involved finding \emph{attenuation factors} $c$ ($0<c<1$) such that (\ref{proportionality}) holds approximately for a range of $\mathbf{x}^T\boldsymbol{\beta}$.

Although these strategies recognize the difference between marginal and conditional models, they fail to provide a single model that both has parameters with a marginal interpretation and allows one to easily make predictions at the individual level.  Instead, they provide a relationship that one could use to obtain parameters with an alternative interpretation after a conditional or marginal model has been fit.  In this paper, we explore a class of conditionally-specified, likelihood-based models with a direct marginal interpretation for the parameters.  We say that a GLMM is \emph{marginally interpretable} if and only if for all $\mathbf{x}^T\boldsymbol{\beta}$  
\vskip -0.2 in
\begin{equation} \label{relationship} 
\int h(\mathbf{x}^T\boldsymbol{\beta} + \mathbf{d}^T\mathbf{u})f_\mathbf{U}(\mathbf{u}) d\mathbf{u} = h(\mathbf{x}^T\boldsymbol{\beta}),
\end{equation}   
where $f_\mathbf{U}$ is the joint density of the random effects.  This property does not hold for most common choices of link function $g(\cdot)$ and random effects density $f_\mathbf{U}$.  In particular, when mean-zero normal random effects are paired with canonical link functions, (\ref{relationship}) is generally not satisfied.

One model that satisfies (\ref{relationship}) is the \emph{marginalized multilevel model} of \citet{heagerty1999} and \citet{heagertyzeger2000}, which expresses the conditional mean as $\mathrm{E}(Y|\mathbf{U})=h(\Delta+\mathbf{d}^T\mathbf{U})$, where $\Delta$ is defined implicitly by the equation
\vskip-0.2in
\begin{equation} \label{mmm}
\int h(\Delta + \mathbf{d}^T\mathbf{u})f_\mathbf{U}(\mathbf{u}) d\mathbf{u} = h(\mathbf{x}^T\boldsymbol{\beta}).
\end{equation}  
In this article, we parameterize $\Delta=\mathbf{x}^T\boldsymbol{\beta} + \mathbf{d}^T\mathbf{a}$, where $\mathbf{d}^T\mathbf{a}$ is known as the \emph{adjustment} and is defined such that (\ref{mmm}) holds.  The adjustment $\mathbf{d}^T\mathbf{a}$ is a function that potentially depends on the fixed portion of the model $\mathbf{x}^T\boldsymbol{\beta}$, the parameters characterizing the random effects distribution $f_\mathbf{U}$, and the random effects design $\mathbf{d}$.  

We view $\mathbf{d}^T\mathbf{a}$ as a location shift of the random effects distribution.  As such, we cease to conceptualize each realization of a random effect as a single value shared by all observations in the same group or cluster.  Rather, observations sharing the same random effect are viewed as having a value representing the same quantile of a location family of distributions.  Since the location of the random effects distribution for a particular observation depends on the covariates for that observation, the value associated with a specific realization of a random effect varies across observations in the same group or cluster.  An example of when different observations with the same random effect could be associated with different values for the random effect is when there are repeated measurements on an individual over time and the covariates vary with time.  Additional details regarding this characterization of random effects can be found in Section~\ref{sec:interpretation}.  

Parameter estimation for GLMMs can be difficult because evaluation of the marginal likelihood often involves an analytically intractable integral.  Thus, one must use numerical integration or employ a method that approximates the likelihood, such as \emph{penalized quasi-likelihood} (PQL) \citep{breslowclayton1993}. An alternative is to adopt a Bayesian framework and employ Markov chain Monte Carlo (MCMC) to produce samples from the posterior for inference \citep[e.g.][]{zegerkarim1991, gamerman1997, migliorettiheagerty2004}.  The adjustment in our proposed model can easily be incorporated into an MCMC algorithm and, although not always available in closed form, can be calculated efficiently.  In addition to discussing the form of the adjustment for many commonly-used GLMMs, we will discuss a novel approach for evaluating the logistic-normal integral that allows for accurate and efficient calculation of the adjustment for a model with a logit link and normal random effects.

In the next section we formally introduce our proposed characterization of the model and discuss properties of the model.  We introduce the concept that replication is required in a mixed model before we can consistently estimate the random effects variance, and show how our parameterization has this property whereas a commonly-specified model fails to have this important property.  Section~\ref{sec:adjustment} provides more detail about the form of the model for specific link functions.  In Section~\ref{sec:computation} we discuss a Bayesian approach to fitting the proposed model, including an algorithm for efficiently computing the logistic-normal integral.  Applications of the proposed model to data from a teratological experiment on rats and from a clinical trial on epileptics are given in Section~\ref{sec:applications}.  Finally, Section~\ref{sec:discussion} contains a discussion of possible avenues for further research.  Proofs of all results, along with other technical details, can be found in the Supplementary Material.

\vspace{-0.2in} \label{sec:model}
\section{Marginally Interpretable Generalized Linear Mixed Models}
\vskip-0.05in
\hspace{.9cm} Formally, we propose the following mixed effects model.  For a response $Y_i$ ($i=1,\dots,N$), a $p$-vector of predictors $\mathbf{x}_i$, and a $q$-vector of random effects $\mathbf{U}_i$, we model the conditional mean as
\vspace{-0.25 in}
\begin{equation} \label{MIGLMM}
\mathrm{E}(Y_i|\mathbf{U}_i=\mathbf{u}) = h(\mathbf{x}_i^T\boldsymbol{\beta} + \mathbf{d}_i^T\mathbf{u} + \mathbf{d}_i^T\mathbf{a}_i), \quad i=1,\dots,N,
\end{equation} 
\vskip-0.05 in \noindent
where the $\mathbf{U}_i$ have joint density given by $f_\mathbf{U}(\mathbf{U}_i)$ and, conditional on the $\mathbf{U}_i$, the $Y_i$ are mutually independent with density $f_{Y|\mathbf{U}}(Y_i|\mathbf{U}_i)$ for each $i$.
The adjustment $\mathbf{d}_i^T\mathbf{a}_i$, when it exists, is defined implicitly by the equation
\vskip-0.3in
\begin{equation} \label{defineadjust}
h(\mathbf{x}_i^T\boldsymbol{\beta}) = \int h(\mathbf{x}_i^T\boldsymbol{\beta} + \mathbf{d}_i^T\mathbf{u} + \mathbf{d}_i^T\mathbf{a}_i) f_\mathbf{U}(\mathbf{u})d\mathbf{u}, 
\end{equation} 
and is included to ensure that the model we consider is marginally interpretable as defined in (\ref{relationship}).  If the random effects all had zero variance, then all effects in the model would be fixed and our expression for the conditional mean would reduce to the marginal mean $h(\mathbf{x}_i^T\boldsymbol{\beta})$.  When the random effects have nonzero variance, the adjustment is required to preserve the marginal mean. 

By including the adjustment in our GLMM, we have specified a formal statistical model for which the parameters $\boldsymbol{\beta}$ have a marginal interpretation.  Likelihood-based methods can be employed to fit this model and the results can be used to make inferences about the marginal mean and to make predictions at the individual level.  The proposed model is superior to a typical marginal model defined only in terms of the mean and covariance structure because it is a fully-specified model with a likelihood.  Further, we argue that the proposed model could be preferred to a conventional GLMM that does not include the adjustment because it provides a direct marginal interpretation of the parameters.  

\vspace{-0.15in}
\subsection{Interpretation of Random Effects} \label{sec:interpretation}
\vskip-0.05in
\hspace{.9cm} Traditionally, one views each realization of a random effect as a single value that applies to all units in a group of observations sharing that random effect.  For example, each random intercept in a conventional random intercepts model corresponds to a shift in the mean response, and for all units with the same random intercept the mean is shifted by the same amount.  When the curvature of the link function is not uniform across the range of the covariates it may be necessary to associate each unit in the same group with a different value of the random effect in order to preserve the marginal mean.  Thus, the idea that all units sharing the same random intercept are shifted by the same amount is not always applicable in a marginally interpretable model.

The role of the adjustment in the proposed model is to ensure that (\ref{relationship}) holds by accounting for the curvature of the inverse link function.  The form of the adjustment is determined by the choice of link function and random effects distribution, whereas its specific value typically depends on $\mathbf{x}_i^T\boldsymbol{\beta}$.  Exceptions for which $\mathbf{d}_i^T\mathbf{a}_i$ does not depend on $\mathbf{x}_i^T\boldsymbol{\beta}$ are models with an identity link or a log link.  For a model with an identity link, (\ref{relationship}) holds as long as $\mathrm{E}(\mathbf{U}_i)=\mathbf{0}$.  Thus, a standard linear mixed model is marginally interpretable without including an adjustment.  Table~\ref{links} summarizes the form and existence of $\mathbf{d}_i^T\mathbf{a}_i$ for several common choices of link function.  More specifically, this table describes the relationship between $\mathbf{x}_i^T\boldsymbol{\beta}$ and $\mathbf{d}_i^T\mathbf{a}_i$ for various link functions and random effects distributions, and also indicates whether or not there exists a closed-form solution for $\mathbf{d}_i^T\mathbf{a}_i$. The interplay between $h(\cdot)$ and $f_\mathbf{U}$ is explored in greater depth in Section~\ref{sec:adjustment}.  

\begin{table}[t]
\caption{\label{links} Form and existence of the adjustment for common link functions}
\begin{center}
\begin{tabular}{c|c|c|c}
Link Function	&Distribution of $\mathbf{U}_i$ &Form of $\mathbf{d}_i^T\mathbf{a}_i$ &Closed Form? \\ \hline
identity	&mean exists and equals zero &zero &yes	 \\ \hline
log	&exponential tails &independent of $\mathbf{x}_i^T\boldsymbol{\beta}$ &yes	 \\ \hline
probit	&Gaussian &linear in $\mathbf{x}_i^T\boldsymbol{\beta}$ &yes  \\
&non-Gaussian &nonlinear in $\mathbf{x}_i^T\boldsymbol{\beta}$ &no  \\ \hline
logit	 &bridge distribution &linear in $\mathbf{x}_i^T\boldsymbol{\beta}$ &yes \\
&most other distributions &nonlinear in $\mathbf{x}_i^T\boldsymbol{\beta}$ &no  \\ \hline
complementary log-log &bridge distribution &linear in $\mathbf{x}_i^T\boldsymbol{\beta}$ &yes \\
&most other distributions &nonlinear in $\mathbf{x}_i^T\boldsymbol{\beta}$ &no  \\ \hline
square root	&restrictions on domain		&nonlinear in $\mathbf{x}_i^T\boldsymbol{\beta}$ &yes \\ \hline
reciprocal	&$\mathrm{E}\{1/(\mathbf{x}_i^T\boldsymbol{\beta} + \mathbf{d}_i^T\mathbf{U}_i)\}$ exists	&see Section~\ref{sec:restrictedlinks} &see Section~\ref{sec:restrictedlinks}
\end{tabular}
\vspace{-0.2in}
\end{center}
\end{table}

Although we write $\mathbf{d}_i^T\mathbf{a}_i$ as a term in the conditional mean, as discussed in Section~\ref{sec:introduction} we view it as part of the random effects distribution.  When $\mathbf{d}_i^T\mathbf{a}_i$ depends on $\mathbf{x}_i^T\boldsymbol{\beta}$, this means that the value associated with the random effect for a particular observation depends on the covariates for that observation.  Thus, each realization of a random effect represents a set of potential values with the value for a specific observation determined by $\mathbf{x}_i^T\boldsymbol{\beta}$.  This differs from the traditional formulation of a random effect and it allows one to separate systematic variation in the population, captured by $\mathbf{x}_i^T\boldsymbol{\beta}$, from individual-level variation, captured by $\mathbf{d}_i^T\mathbf{U}_i + \mathbf{d}_i^T\mathbf{a}_i$. \\

A situation where this new formulation of a random effect might arise is in a multilevel model where the covariates $\mathbf{x}_i$ differ across individual units in the same group or cluster.  For example, students within the same class are liable to have different characteristics.  Depending on the choice of link function, the adjustment could be different for different units in the same group.  Thus, within a single group, the shift in the mean response associated with the random effect for that group could vary with the measured covariates for the individual units sharing that random effect.

\vspace{-0.15in}
\subsection{Consistent Estimation of the Random Effects Variance} \label{sec:consistency}
\vskip-0.05in
\hspace{.9cm} The random effects in a mixed model provide a means of introducing dependence and overdispersion into the model.  For example, repeated measures on a subject may be systematically large, or count data may show extra-Poisson variation.  In simple models, the degree of dependence/overdispersion is determined by the distribution of the random effects, most typically by its variance.  Intuitively, one must have replication to consistently estimate the random effects variance.  Consider, for example, the following hierarchical model:  
\vskip -0.55 in   
\begin{eqnarray*}
Z_i & \sim & F_{\sigma^2}; \\
Y_i | Z_i=z_i & \sim & \mathrm{Bernoulli}(z_i),
\end{eqnarray*}
\vskip -0.2 in \noindent
where $i=1,\dots,N$ and $F_{\sigma^2}$ is an arbitrary distribution on $(0,1)$ with mean $\mu$ and variance $\sigma^2$.  To obtain the marginal model, one must integrate over $Z_i$.  Irrespective of $\sigma^2$, the resulting marginal distribution for $Y_i$ is $\mathrm{Bernoulli}(\mu)$.  No matter how many of these Bernoullis are collected, there is no replication tied to a single random effect, no information is obtained about $\sigma^2$, and hence $\sigma^2$ cannot be estimated consistently. 

Although it seems natural that consistent estimation of the random effects variance requires replication, \citet{kimkim2011} proved a surprising result for a conventional Bernoulli GLMM.  Namely, they showed that the maximum likelihood estimator is strongly consistent for the random effects variance $\sigma^2$, even without replication. 
We call this the Kim Paradox, stating a slightly different result than is presented in \citet{kimkim2011} and placing the result in our notation.  

\vskip 0.15 in
\noindent
{\bf The Kim Paradox:} 
\emph{With no replication, one can estimate $\sigma^2$ consistently.}  Let parameters $\beta_0$, $\beta_1 \neq 0$, and $\tau^2 > 0$ be fixed and known, and let $X_i \sim \mathrm{N}(0,\tau^2)$ and $U_i \sim \mathrm{Uniform}(-c,c)$ ($c>0$, $i=1,2,\dots$), be independent sequences of random variables.  Furthermore, let $g(\cdot) = \mathrm{logit}(\cdot)$, and define the conditionally independent sequence $Y_i | X_i=x_i, U_i=u_i \sim \mathrm{Bernoulli}\{h(\beta_0 + \beta_1 x_i + u_i)\}$ for $i = 1, 2, \dots$.  Then $\widehat{\sigma}^2$, the maximum likelihood estimator of $\sigma^2$, is consistent.  

\vskip 0.15 in
The Kim Paradox arises from the fact that $\mathrm{E}[h(\beta_0 + \beta_1 x_i + U_i)]$, the expectation of the conditional mean in a conventional GLMM, is distorted by the random effects in such fashion that there is a $1-1$ mapping between $\sigma^2$ and $\mathrm{E}[h(\beta_0 + \beta_1 x_i + U_i)]$.  This $1-1$ mapping, along with a rich enough set of $x_i$, ensures that the marginal mean functions are identifiable, and consistency of $\sigma^2$ follows.  A marginally interpretable Bernoulli GLMM of the form (\ref{MIGLMM}) and (\ref{defineadjust}) resolves the Kim Paradox because the marginal mean is unaffected by changes in $\sigma^2$.  Consequently, the data contain {\em no} information about $\sigma^2$, and one cannot obtain a consistent estimator of the random effects variance without replication.  This is stated more formally in the following proposition:  

\begin{prop} \label{bernoullilike} 
If $Y_i|U_i$ is Bernoulli-distributed and we have a marginally interpretable GLMM of the form given by (\ref{MIGLMM}) and (\ref{defineadjust}) for which the random intercepts $U_i$, $i=1,\dots,N$, are independently distributed, then the marginal density of $Y_i$ does not depend in any way on the distribution of $U_i$.
\end{prop}

\vspace{-0.23in}
\section{The Form of the Adjustment} \label{sec:adjustment}
\vskip-0.07in
\hspace{.9cm} As discussed in Section~\ref{sec:interpretation} and summarized in Table~\ref{links}, the form of $\mathbf{d}_i^T\mathbf{a}_i$ depends on the choice of link function and random effects distribution.  This section provides more detail about the form of $\mathbf{d}_i^T\mathbf{a}_i$ for several common choices of link function.

\vspace{-0.18in}
\subsection{Log Link} \label{sec:loglink}
\vskip-0.07in
\hspace{.9cm} Consider a GLMM with a log link.  That is, let the link function be $g(\cdot)=\log(\cdot)$ with the inverse link $h(\cdot)=\exp(\cdot)$.  In this case, $\mathbf{d}_i^T\mathbf{a}_i$ is defined such that 
\vskip-0.2in
\begin{equation} \label{logrelation}
\exp(\mathbf{x}_i^T\boldsymbol{\beta})=\int \exp(\mathbf{x}_i^T\boldsymbol{\beta}+\mathbf{d}_i^T\mathbf{u}+ \mathbf{d}_i^T\mathbf{a}_i)f_\mathbf{U}(\mathbf{u})d\mathbf{u}.  
\end{equation}   
Solving (\ref{logrelation}) for $\mathbf{d}_i^T\mathbf{a}_i$ leads to the following proposition:
\begin{prop} \label{logadjust} 
For $h(\cdot)=\exp(\cdot)$, a model of the form given by (\ref{MIGLMM}) and (\ref{defineadjust}) is marginally interpretable if and only if $\mathbf{d}_i^T\mathbf{a}_i = -\log\{M_\mathbf{U}(\mathbf{d}_i)\}$, where $M_{\mathbf{U}}(\mathbf{d}_i)=\mathrm{E}\{\exp(\mathbf{d}_i^T\mathbf{U}_i)\}$ is the moment-generating function of $\mathbf{U}_i$ evaluated at $\mathbf{d}_i$. 
\end{prop}
\noindent From Proposition~\ref{logadjust} we obtain the following corollary:
\begin{corol} \label{logexist} 
For a GLMM with inverse link function $h(\cdot)=\exp(\cdot)$, an adjustment $\mathbf{d}_i^T\mathbf{a}_i$ that makes the model marginally interpretable exists if and only if $M_{\mathbf{U}}(\mathbf{d}_i)$ exists.
\end{corol}
\noindent These results constrain the set of possible random effects distributions that can be used with this model to those with exponential tails.  Consequently, the t-distribution is not a valid random effects distribution for a marginally interpretable GLMM with a log link.  One could, however, use a mixture of normal distributions to approximate a t-distribution.

To better understand the role of the adjustment for a model with a log link, consider the case of a single random intercept $U_i \sim \mathrm{N}(0,\sigma^2)$.  In this case, $d_i=1$ for all $i=1,\dots,N$, and the adjustment $d_ia_i$ is expressed simply as $a_i$.  From Proposition~\ref{logadjust} we have the following result:
\begin{corol} \label{lognormal}
A model of the form $\mathrm{E}(Y_i|U_i=u) = \exp(\mathbf{x}_i^T\boldsymbol{\beta} + u + a_i)$ for which $U_i \sim \mathrm{N}(0,\sigma^2)$ is marginally interpretable if and only if $a_i=-\sigma^2/2$ for all $i=1,\dots,N$.
\end{corol}
\noindent In this situation, the adjustment depends only on the random effects variance $\sigma^2$ and is independent of $\mathbf{x}_i^T\boldsymbol{\beta}$.  It is simply an additive offset on the log scale that pulls the conditional mean $\mathrm{E}(Y_i|U_i)$ down by the same amount for all $i=1,\dots,N$.  This effectively shifts the location of the random effects distribution in a manner that makes the model marginally interpretable.  Since the inverse link function $\exp(\cdot)$ is convex, when the random effect has mean zero and no adjustment is made we know by Jensen's inequality that
\vspace{-0.2in}
\begin{equation*} \label{jensens}
\mathrm{E}(Y_i) = \mathrm{E}\{\mathrm{E}(Y_i|U_i)\} = \mathrm{E}\{\exp(\mathbf{x}_i^T\boldsymbol{\beta}+U_i)\} \ge \exp\{\mathrm{E}(\mathbf{x}_i^T\boldsymbol{\beta}+U_i)\} = \exp(\mathbf{x}_i^T\boldsymbol{\beta}).
\end{equation*} 
\vskip-0.2in \noindent
By pulling the conditional mean down, the adjustment counteracts the convexity of the inverse link function so that the marginal mean $\mathrm{E}(Y_i)$ is equal to $\exp(\mathbf{x}_i^T\boldsymbol{\beta})$.  

\vspace{-0.15in}
\subsection{Links with Bounded Domain} \label{sec:boundedlinks}
\vskip-0.05in
\hspace{.9cm} Several common link functions, including the probit, logit, and complementary log-log, are defined only on a bounded subset of the real line.  In turn, the range of the corresponding inverse link function $h(\cdot)$ is constrained to a bounded interval.  For models with such a link function, the following theorem applies:   
\begin{thrm} \label{boundedh} 
Consider a model of the form given in (\ref{MIGLMM}) with $h: \mathbb{R} \rightarrow \mathcal{I}=[\ell,u]$.  Suppose $h(\cdot)$ is increasing and continuous, with $h(\eta) \rightarrow \ell$ as $\eta \rightarrow -\infty$ and $h(\eta) \rightarrow u$ as $\eta \rightarrow \infty$.  Then an adjustment $\mathbf{d}_i^T\mathbf{a}_i$ that satisfies (\ref{defineadjust}) exists for any choice of random effects distribution.
\end{thrm}
\noindent Thus, one can always construct a model to be marginally interpretable when using a link function defined only on a bounded interval.  We now discuss link functions with this property.

\vspace{-0.15in}
\subsubsection*{Probit Link} \label{sec:probitlink}
\vskip-0.05in
\hspace{.9cm} Let $g(\cdot)=\Phi^{-1}(\cdot)$ and $h(\cdot)=\Phi(\cdot)$, where $\Phi(\cdot)$ is the cumulative distribution function of a standard normal distribution.  The range of the inverse link function $h(\cdot)$ is the bounded interval $(0,1)$.  Therefore, Theorem~\ref{boundedh} applies  and an adjustment $\mathbf{d}_i^T\mathbf{a}_i$ that makes the model marginally interpretable exists regardless of the choice of random effects distribution.

For a model with a probit link and normal random effects the adjustment has a closed form.  Specifically, let $\mathbf{U}_i \sim \mathrm{N}_q(\mathbf{0},\boldsymbol{\Sigma})$, where $\boldsymbol{\Sigma}$ is a covariance matrix.  One can show that without the adjustment this model would satisfy a multivariate analogue to (\ref{proportionality}) with
$c=(1+\mathbf{d}_i^T\boldsymbol{\Sigma}\mathbf{d}_i)^{-1/2}$ \citep[see][]{mccullochetal2008}.  This leads to the following proposition:
\begin{prop} \label{probitadjust} 
For $h(\cdot)=\Phi(\cdot)$ and $\mathbf{U}_i \sim \mathrm{N}_q(\mathbf{0},\boldsymbol{\Sigma})$, a model of the form given by (\ref{MIGLMM}) and (\ref{defineadjust})
is marginally interpretable if and only if $\mathbf{d}_i^T\mathbf{a}_i = \{(1+\mathbf{d}_i^T\boldsymbol{\Sigma}\mathbf{d}_i)^{1/2}-1\} \mathbf{x}_i^T\boldsymbol{\beta}$. 
\end{prop}
\noindent Thus, $\mathbf{d}_i^T\mathbf{a}_i$ is a linear function of $\mathbf{x}_i^T\boldsymbol{\beta}$.  This linearity is beneficial for model interpretation, but models with a probit link are generally difficult to interpret because the probit does not have the convenient log-odds interpretation of the logit.  We therefore focus on models with a logit link. 

\vspace{-0.15in}
\subsubsection*{Logit Link} \label{sec:logitlink}
\vskip-0.1in
\hspace{.9cm} Consider a GLMM with link function $g(\mu)=\log\{\mu/(1-\mu)\}$ and inverse link function $h(\eta)=\exp(\eta)/\{1+\exp(\eta)\}=1/\{1+\exp(-\eta)\}$.  This function $g(\cdot)$ is known both as the \emph{logit link} and as the \emph{logistic link}.  The adjustment $\mathbf{d}_i^T\mathbf{a}_i$ for this model is defined such that 
\vskip-0.2in
\begin{equation} \label{logitadjust}
\frac{1}{1 + e^{-\mathbf{x}_i^T\boldsymbol{\beta}}} = 
\int \frac{1}{1 + e^{-(\mathbf{x}_i^T\boldsymbol{\beta} + \mathbf{d}_i^T\mathbf{u} + \mathbf{d}_i^T\mathbf{a}_i)}} f_\mathbf{U}(\mathbf{u}) d\mathbf{u}. 
\end{equation}   
Once again, the range of the inverse link function $h(\cdot)$ is the bounded interval $(0,1)$.  Thus, by Theorem~\ref{boundedh}, there are no restrictions on the choice of the random effects distribution.  However, for most choices of random effects distribution the integral on the right-hand-side of (\ref{logitadjust}) is analytically intractable and there is no closed-form solution for $\mathbf{d}_i^T\mathbf{a}_i$.  One exception is the \emph{bridge distribution} derived by \citet{wanglouis2003}.  Provided the model contains just a single random intercept, the bridge distribution leads to a closed-form solution for $\mathbf{d}_i^T\mathbf{a}_i$ that is linear as a function of $\mathbf{x}_i^T\boldsymbol{\beta}$.

It is most common to assume that $f_\mathbf{U}$ is a normal density.  In this article we develop a novel and efficient method for calculating  $\mathbf{d}_i^T\mathbf{a}_i$ under this assumption for the random effects distribution.  Our algorithm for evaluating the logistic-normal integral exploits a recursive formula developed by \citet{pirjol2013} that provides an exact solution to the logistic-normal integral on a specifically-defined, evenly-spaced grid.  \citet{pirjol2013} demonstrated that the integral 
\vskip -0.2 in
\begin{equation} \label{pirjolintegral}
\varphi(\mu,\sigma^2) = \int \frac{1}{1+e^w} \frac{1}{\sqrt{2\pi\sigma^2}} \exp\bigg{\{}-\frac{1}{2\sigma^2} (w-\mu)^2\bigg{\}}dw 
\end{equation} 
satisfies the recursion 
\vskip -0.4 in
\begin{equation} \label{recursion}  
\varphi(\mu+\sigma^2,\sigma^2) = e^{-\mu-\frac{\sigma^2}{2}} \{1-\varphi(\mu,\sigma^2)\},
\end{equation} 
\vskip -0.1 in \noindent
where $\varphi(0,\sigma^2)=1/2$.  See Section~\ref{sec:logitnormal} for details on how this result is used to compute $\mathbf{d}_i^T\mathbf{a}_i$ accurately and efficiently.  

\begin{figure}[t]
\begin{center}
\includegraphics[width=4.2in]{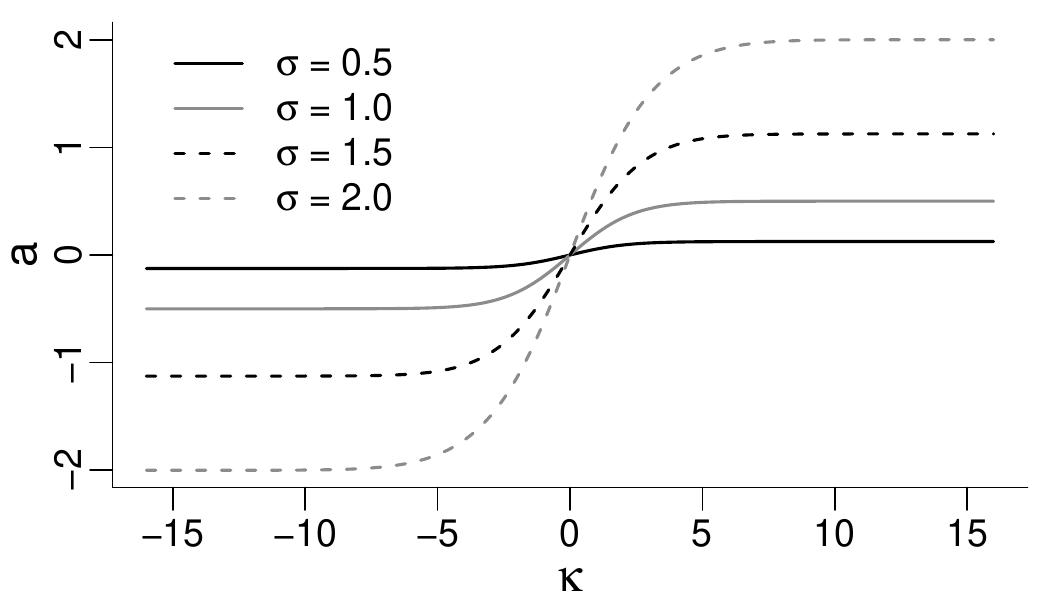}
\end{center}
\vspace{-0.3in}
\caption{\label{adjusts} Plot of the adjustment $a$ as a function of $\kappa$ for various values of $\sigma$, where $\kappa=\mathbf{x}_i^T\boldsymbol{\beta}$ represents the fixed effects portion of the model.} 
\vspace{-0.05in}
\end{figure}

For a model with a logit link, whether $f_\mathbf{U}$ is assumed to be normal or not, both the direction and magnitude of the adjustment depend on $\mathbf{x}_i^T\boldsymbol{\beta}$.  In fact, the adjustment $\mathbf{d}_i^T\mathbf{a}_i$ is typically a nonlinear function of $\mathbf{x}_i^T\boldsymbol{\beta}$.  This is illustrated in Figure~\ref{adjusts} for the case of a single normal random intercept $U_i \sim \mathrm{N}(0,\sigma^2)$.  In light of the forthcoming Proposition~\ref{dimreduce}, the same picture would apply for $q$ normal random effects $\mathbf{U}_i \sim \mathrm{N}_q(\mathbf{0},\boldsymbol{\Sigma})$ if we were to replace $\sigma^2$ with $\mathbf{d}_i^T\boldsymbol{\Sigma}\mathbf{d}_i$.  The direction of the adjustment is driven by the convexity of the inverse link function.  The function $h(\eta)$ is convex for $\eta<0$ and concave for $\eta>0$.  Hence, the adjustment is negative when $\mathbf{x}_i^T\boldsymbol{\beta}<0$ and positive when $\mathbf{x}_i^T\boldsymbol{\beta}>0$.  It is also evident from Figure~\ref{adjusts} that the magnitude of $\mathbf{d}_i^T\mathbf{a}_i$ is increasing in both $\sigma^2$ and $|\mathbf{x}_i^T\boldsymbol{\beta}|$.  For very large $\mathbf{x}_i^T\boldsymbol{\beta}$ we have the following result:

\begin{prop} \label{logitlimit} 
For $h(\cdot)=\mathrm{logit}^{-1}(\cdot)$ and $\mathbf{U}_i \sim \mathrm{N}_q(\mathbf{0},\boldsymbol{\Sigma})$, the value of $\mathbf{d}_i^T\mathbf{a}_i$ that allows a model of the form given by (\ref{MIGLMM}) to satisfy (\ref{defineadjust}) converges to $\frac{1}{2}\mathbf{d}_i^T\boldsymbol{\Sigma}\mathbf{d}_i \times \mathrm{sign}(\mathbf{x}_i^T\boldsymbol{\beta})$ as $|\mathbf{x}_i^T\boldsymbol{\beta}| \rightarrow \infty$.
\end{prop}

Figure~\ref{adjusts} also shows that, for a model with a logit link, observations that have different values of the covariates $\mathbf{x}_i$ also have different adjustments.  This helps illustrate the point from Section~\ref{sec:interpretation} that with the logit link, units that share a random effect but have different measured covariates do not have their means shifted by the same amount.  Rather, the magnitude of the shift associated with the random effect for each observation is dependent on the value of $\mathbf{x}_i^T\boldsymbol{\beta}$ for that observation. 

\vspace{-0.15in}
\subsubsection*{Complementary Log-Log Link} \label{sec:clogloglink}
\vskip -0.05 in
\hspace{.9cm} Consider a GLMM with $g(\mu)=\log\{-\log(1-\mu)\}$ and $h(\eta)=1-\exp\{-\exp(\eta)\}$.  The adjustment $\mathbf{d}_i^T\mathbf{a}_i$ for this model is defined such that 
\vskip-0.2in
\begin{equation*} \label{cloglogadjust}
\exp\{-\exp(\mathbf{x}_i^T\boldsymbol{\beta})\} = 
\int \exp\{-\exp(\mathbf{x}_i^T\boldsymbol{\beta} + \mathbf{d}_i^T\mathbf{u} + \mathbf{d}_i^T\mathbf{a}_i)\} f_\mathbf{U}(\mathbf{u}) d\mathbf{u}. 
\end{equation*} 
\noindent As with a logit link, there are no restrictions on the choice of the random effects distribution, but in most cases there is no closed-form solution for $\mathbf{d}_i^T\mathbf{a}_i$.  \citet{wanglouis2003} also derived a bridge distribution for this link function that leads to a closed-form adjustment that is linear in $\mathbf{x}_i^T\boldsymbol{\beta}$.  For more conventional choices of random effects distribution, namely normal random effects, one must use some form of numerical integration, such as Gaussian quadrature, to calculate $\mathbf{d}_i^T\mathbf{a}_i$. 

\vspace{-0.15in}
\subsection{Links with Range Restrictions} \label{sec:restrictedlinks}
\vskip -0.05 in
\hspace{.9cm} A number of common link functions map into a proper subset of the real line and therefore require conditions on $\mathbf{x}_i^T\boldsymbol{\beta}$ to ensure that the model is defined.  For example, the square root transformation is typically defined to have nonnegative range, and no real number has a reciprocal of zero. Additive random effects with support on the entire real line could lead to problems in models with these link functions because $\mathbf{x}_i^T\boldsymbol{\beta} + \mathbf{d}_i^T\mathbf{U}_i$ could fall outside the domain of the inverse link function $h(\cdot)$.  Thus, special care must be taken with these link functions, as described below.

\vspace{-0.15in}
\subsubsection*{Square Root Link} \label{sec:sqrtlink}
\vskip -0.05 in
\hspace{.9cm} Consider a GLMM with link function $g(\mu)=\mu^{1/2}$ and inverse link function $h(\eta)=\eta^2$.  For such a model one typically includes the restriction that $\mathbf{x}_i^T\boldsymbol{\beta} \ge 0$.  Including the adjustment, we adopt the restriction that $\mathbf{x}_i^T\boldsymbol{\beta} + \mathbf{d}_i^T\mathbf{a}_i \ge 0$. The adjustment is defined such that 
\vskip-0.2in
\begin{equation} \label{sqrtadjustdef}
(\mathbf{x}_i^T\boldsymbol{\beta})^2 = 
\int (\mathbf{x}_i^T\boldsymbol{\beta} + \mathbf{d}_i^T\mathbf{u} + \mathbf{d}_i^T\mathbf{a}_i)^2 f_\mathbf{U}(\mathbf{u}) d\mathbf{u}. 
\end{equation}   
If we assume $\mathrm{E}(\mathbf{U}_i)=\mathbf{0}$, then (\ref{sqrtadjustdef}) reduces to 
\vskip -0.2 in
\begin{equation} \label{sqrtadjustquad}
(\mathbf{x}_i^T\boldsymbol{\beta})^2 = 
(\mathbf{x}_i^T\boldsymbol{\beta} + \mathbf{d}_i^T\mathbf{a}_i)^2 + \mathrm{Var}(\mathbf{d}_i^T\mathbf{U}_i), 
\end{equation}   
which is quadratic in $\mathbf{d}_i^T\mathbf{a}_i$ and leads to the following result:

\begin{prop} \label{sqrtadjust}
For $h(\eta)=\eta^2$ \ and \  $\mathrm{E}(\mathbf{U}_i)=\mathbf{0}$, a model of the form given by (\ref{MIGLMM}) and (\ref{defineadjust}) subject to the restriction that $\mathbf{x}_i^T\boldsymbol{\beta} + \mathbf{d}_i^T\mathbf{a}_i \ge 0 $ is a marginally interpretable model if and only if
$\mathbf{d}_i^T\mathbf{a}_i = -\mathbf{x}_i^T\boldsymbol{\beta} + \{(\mathbf{x}_i^T\boldsymbol{\beta})^2 - \mathrm{Var}(\mathbf{d}_i^T\mathbf{U}_i)\}^{1/2}$.
\end{prop}   
\noindent Thus, for a model with a square root link, $\mathbf{d}_i^T\mathbf{a}_i$ is a nonlinear function of $\mathbf{x}_i^T\boldsymbol{\beta}$ and is only defined when $\mathbf{x}_i^T\boldsymbol{\beta} \ge \{\mathrm{Var}(\mathbf{d}_i^T\mathbf{U}_i)\}^{1/2}$.  If the random effects variance is too large, the model cannot be fit.

\vspace{-0.15in}
\subsubsection*{Reciprocal Link} \label{sec:reciprocallink}
\vskip -0.05 in
\hspace{.9cm} Finally, consider a model for which $g(\mu)=h(\mu)=1/\mu$.  For a fixed effects model with this link one typically includes the restriction that $\mathbf{x}_i^T\boldsymbol{\beta} > 0$.  When a random intercept $U_i$ is included in the model, the fact that $h(\cdot)$ tends to infinity as its argument approaches zero forces us to also include restrictions on the distribution of $U_i$.  In particular, we want a model for which
\vskip -0.2 in
\begin{equation} \label{recipadjust}
\frac{1}{\mathbf{x}_i^T\boldsymbol{\beta}} = \int \frac{1}{ \mathbf{x}_i^T\boldsymbol{\beta} + u}f_U(u) du.
\end{equation}   
Therefore, $f_U$ must be defined such that the integral on the right-hand-side of (\ref{recipadjust}) exists.  This restriction forces us to move away from models of the form given in (\ref{MIGLMM}).  Rather than adjusting the location of the random effect based on each individual's observed covariates, we must alter the shape of the distribution of the random effect based on the observed covariates.

One distribution for $U_i$ that allows us to satisfy (\ref{recipadjust}) is a shifted gamma distribution.  Specifically, let $\mathbf{x}_i^T\boldsymbol{\beta} + U_i$ follow a gamma distribution with shape parameter $\alpha_i$ and rate parameter $\beta_i$ so that $\mathrm{E}(U_i)=\alpha_i\beta_i-\mathbf{x}_i^T\boldsymbol{\beta}$.  Then the integral on the right-hand-side of (\ref{recipadjust}) is equal to $\{\beta_i(\alpha_i-1)\}^{-1}$, and $\mathbf{x}_i^T\boldsymbol{\beta}=\beta_i(\alpha_i-1)$.  By placing additional conditions on $\alpha_i$ and $\beta_i$ one can determine the appropriate gamma distribution for $U_i$ for each $\mathbf{x}_i^T\boldsymbol{\beta}$.  Alternatively, one could let $\mathbf{x}_i^T\boldsymbol{\beta} + U_i$ follow an inverse gamma distribution with parameters $\alpha_i$ and $\beta_i$, and be constrained by the relationship $\mathbf{x}_i^T\boldsymbol{\beta}=(\alpha_i\beta_i)^{-1}$.  In either case it is the shape, not the location, of the random effects distribution that varies with $\mathbf{x}_i^T\boldsymbol{\beta}$ in this marginally interpretable model.

\vspace{-0.15in}
\section{Computation} \label{sec:computation}
\vskip -0.05 in
\hspace{.9cm} Since the adjustment that we propose is a deterministic function of $\mathbf{x}_i^T\boldsymbol{\beta}$ and the parameters characterizing $f_\mathbf{U}$ (which we denote $\boldsymbol{\alpha}$), it can easily be incorporated into techniques that are commonly used to fit conventional GLMMs.  Taking a Bayesian approach, for example, one might employ a Metropolis-Hastings MCMC algorithm to generate samples from the posterior distribution of the unknown parameters.  Such an algorithm involves iteratively proposing values for the unknown quantities $\boldsymbol{\beta}$, $\boldsymbol{\alpha}$, and $\mathbf{U}$ and choosing to either accept or reject those values in an effort to sample from the posterior density $\pi(\boldsymbol{\beta},\boldsymbol{\alpha},\mathbf{U}|\mathbf{Y})$.  Denoting $\boldsymbol{\theta}=(\boldsymbol{\beta},\boldsymbol{\alpha},\mathbf{U})^T$, the acceptance probability at the $t^{th}$ iteration is given by
\vskip -0.2 in
\begin{equation} \label{generalMH}
\mathrm{min}\bigg{\{}1, \frac{f_{\mathbf{Y}|\boldsymbol{\theta}}(\mathbf{Y}|\boldsymbol{\theta}^*)\pi_{\boldsymbol{\theta}}(\boldsymbol{\theta}^*)q(\boldsymbol{\theta}^{(t)}|\boldsymbol{\theta}^*)}{f_{\mathbf{Y}|\boldsymbol{\theta}}(\mathbf{Y}|\boldsymbol{\theta}^{(t)})\pi_{\boldsymbol{\theta}}(\boldsymbol{\theta}^{(t)})q(\boldsymbol{\theta}^*|\boldsymbol{\theta}^{(t)})}\bigg{\}},
\end{equation}
where $\boldsymbol{\theta}^{(t)}$ is the current state of $\boldsymbol{\theta}$, $\boldsymbol{\theta}^*$ is the proposed state of $\boldsymbol{\theta}$, $f_{\mathbf{Y}|\boldsymbol{\theta}}(\cdot)$ is the conditional density of $\mathbf{Y}$, $\pi_{\boldsymbol{\theta}}(\cdot)$ is the prior density for $\boldsymbol{\theta}$, and $q(\cdot)$ is the proposal density.  Within each update of $\boldsymbol{\theta}$ the proposals $\boldsymbol{\beta}^*$ and $\boldsymbol{\alpha}^*$ can be used along with $\mathbf{x}_i$ and $\mathbf{d}_i$, which are treated as fixed and known, to compute the adjustment $\mathbf{d}_i^T\mathbf{a}_i$, which can in turn be included in evaluation of the likelihood. 

The computational expense added to the algorithm by including the adjustment is driven by how difficult it is to compute $\mathbf{d}_i^T\mathbf{a}_i$.  The difficulty varies based on the choice of link function and random effects distribution, and is greater in situations lacking a closed-form solution for $\mathbf{d}_i^T\mathbf{a}_i$.  The most common situation without a closed-form solution is a GLMM with a logit link and normal random effects.  An efficient approach for computing the adjustment in such a case is described below.  This approach relies on the following result, which applies to any GLMM with multivariate normal random effects and allows one to simplify computation by reducing a $q$-dimensional integral to a univariate one:  
\begin{prop} \label{dimreduce} 
For the case when $\mathbf{U}_i \sim \mathrm{N}_q(\mathbf{0},\boldsymbol{\Sigma})$, if the $q$-dimensional integral 
\vskip -0.15 in
\begin{equation*} \label{multinormint}
\int_{\mathbb{R}^q} h(\kappa+\mathbf{d}^T\mathbf{u}+a)\Big{(}\frac{1}{2\pi}\Big{)}^{\frac{q}{2}}|\boldsymbol{\Sigma}|^{-1/2}\exp\!\Big{(}-\frac{1}{2}\mathbf{u}^T\boldsymbol{\Sigma}^{-1}\mathbf{u}\Big{)} d\mathbf{u} 
\end{equation*} 
exists, then it can be expressed as a univariate integral of the form
\vskip -0.15 in
\begin{equation*} \label{univnormint}
\int_{\mathbb{R}} h(\kappa+v+a)\frac{1}{\sqrt{2\pi\tau^2}}\exp\!\bigg{(}-\frac{1}{2\tau^2}v^2\bigg{)}dv.
\end{equation*}
\end{prop}

\vspace{-0.15in}
\subsection{Efficient and Accurate Evaluation of the Logistic-Normal Integral} \label{sec:logitnormal}
\vskip-0.07in
\hspace{.9cm} Although the logistic-normal integral is analytically intractable, several numerical approaches exist for evaluating it.  Common approaches include Gauss-Hermite quadrature and adaptive quadrature schemes.  There also exist a number of algorithms tailored specifically to evaluating the logistic-normal integral, including methods proposed by \citet{crouchspiegelman1990} and \citet{monahanstefanski1992}.  The method of \citet{monahanstefanski1992} involves approximating the inverse logit function $h(z)$ with a weighted mixture of normal distributions
\vskip-0.2in
\begin{equation*} \label{monahanapproxh}
h_k^*(z) = \sum_{i=1}^k p_{k,i}\Phi(zs_{k,i}),
\end{equation*}
where $\Phi(\cdot)$ is the cumulative distribution function of a standard normal distribution and the weights $p_{k,i}$ and $s_{k,i}$ are chosen to minimize the maximum approximation error over all values of $z$.  This leads to to the integral approximation
\vspace{-0.16in}
\begin{equation} \label{monahanapproxint}
\int h(z) \frac{1}{\sigma}\phi\bigg{(}\frac{z-\xi}{\sigma}\bigg{)}dz \approx \int h_k^*(z) \frac{1}{\sigma}\phi\bigg{(}\frac{z-\xi}{\sigma}\bigg{)}dz = \sum_{i=1}^k p_{k,i}\Phi\Bigg{\{}\frac{\xi s_{k,i}}{(1+\sigma^2 s_{k,i}^2)^{1/2}}\Bigg{\}},
\end{equation}
\vskip-0.16in \noindent
which is within $2.1 \times 10^{-9}$ of the true value of the integral for all values of $\kappa$ and $\sigma$ when one uses $k=8$ mixture weights.  One could use fewer than eight mixture weights to improve computational efficiency, but the increase in speed from using fewer weights is small relative to the corresponding loss of accuracy.  We therefore recommend using $k=8$. 

More recently, \citet{pirjol2013} developed the recursive formula given in (\ref{recursion}), which provides an exact solution to the logistic-normal integral on a specific evenly-spaced grid.  To simplify notation we denote $\kappa=\mathbf{x}_i^T\boldsymbol{\beta}$ and $a=a_i$, and use $\varphi(\cdot,\cdot)$ as defined in (\ref{pirjolintegral}).  Given $\kappa$ and $\sigma^2$, the equation we must solve for $a$ when $g(\cdot)=\mathrm{logit}(\cdot)$ and $U_i \sim \mathrm{N}(0,\sigma^2)$ is 
\vspace{-0.25in} 
\begin{equation} \label{logitnormaleqn}
h(\kappa) = 1-\varphi(\kappa+a,\sigma^2).
\end{equation} 
\vskip-0.25in \noindent
Without loss of generality, due to the symmetry of the problem, we need only consider the case of $\kappa > 0$.  When $\kappa < 0$ the adjustment has the same magnitude but opposite sign as if $\kappa=|\kappa|$.  Further, as a consequence of Proposition~\ref{dimreduce}, the strategy described here for the univariate case also applies when the model includes multivariate normal random effects.

Several different techniques, such as binary segmentation or a Newton-Raphson algorithm, can be used to solve (\ref{logitnormaleqn}) for $a$. Any such technique requires evaluating $\varphi(\kappa+a^*,\sigma^2)$ for several potential values $a^*$ of the adjustment $a$.  While we can use (\ref{recursion}) to calculate $\varphi(t\sigma^2,\sigma^2)$ exactly for any integer $t$, it is unlikely that the desired $a$ will be such that $\kappa+a$ is an integer multiple of $\sigma^2$.  Thus, we have need of an approximate numerical integration procedure.  Since the function $\varphi(\cdot,\cdot)$ is decreasing in its first argument, we can use (\ref{recursion}) to quickly identify an interval of length $\sigma^2$ in which $\kappa+a$ must reside.  We denote this interval $(t^*\sigma^2,(t^*+1)\sigma^2)$, where $t^*$ is a nonnegative integer.  By narrowing our search for the correct value of $a$ to such an interval we cut down the required number of evaluations of $\varphi(\kappa+a^*,\sigma^2)$.  

We also use the recursive result of \citet{pirjol2013} to improve the accuracy of the necessary integral approximations.  After identifying
$t^*$, we employ binary segmentation to search within the interval $(t^*\sigma^2,(t^*+1)\sigma^2)$ for the value of $\kappa+a$ satisfying (\ref{logitnormaleqn}).  To evaluate $\varphi(\kappa+a^*,\sigma^2)$ for $\kappa+a^* \in (t^*\sigma^2,(t^*+1)\sigma^2)$ we could simply use (\ref{monahanapproxint}).  However, we find that the Monahan-Stefanski approximation is generally more accurate near zero than away from it.  We therefore use (\ref{monahanapproxint}) to compute $\varphi(\kappa+a^*-t^*\sigma^2,\sigma^2)$ and then apply (\ref{recursion}) $t^*$ times to obtain $\varphi(\kappa+a^*,\sigma^2)$.  

To assess the speed and accuracy of our approach we compared it to both 30-point Gauss-Hermite quadrature and to a direct application of (\ref{monahanapproxint}).  Specifically, for each of the 80 values of $\sigma$ in the set $\{0.05,0.10,\dots,4.00\}$ we evaluated the integral $1-\varphi(\mu,\sigma^2)$ for 1,000 values of $\mu$ in each of the four intervals $[0,\sigma^2]$, $[\sigma^2,2\sigma^2]$, $[2\sigma^2,3\sigma^2]$, and $[3\sigma^2,4\sigma^2]$ using our method, the method of Monahan and Stefanski, 30-point Gauss-Hermite quadrature, and 1,000-point Gauss-Hermite quadrature.  This required 4,000 integral evaluations for each of the 80 values of $\sigma$ and each method.  These evaluations were completed on a Dual Quad Core Xeon 2.66 E5430 computer with 32 gigabytes of RAM.  To ensure a fair comparison of speed, all four approaches were implemented using the Rcpp package in R \citep{rbase,rcpparticle,rcppbook}.  Gauss-Hermite quadrature with 1,000 quadrature points was treated as the \emph{gold standard} to which the other three methods were compared to assess accuracy.   

\begin{figure}[t]
\begin{center}
\includegraphics[width=3.1in]{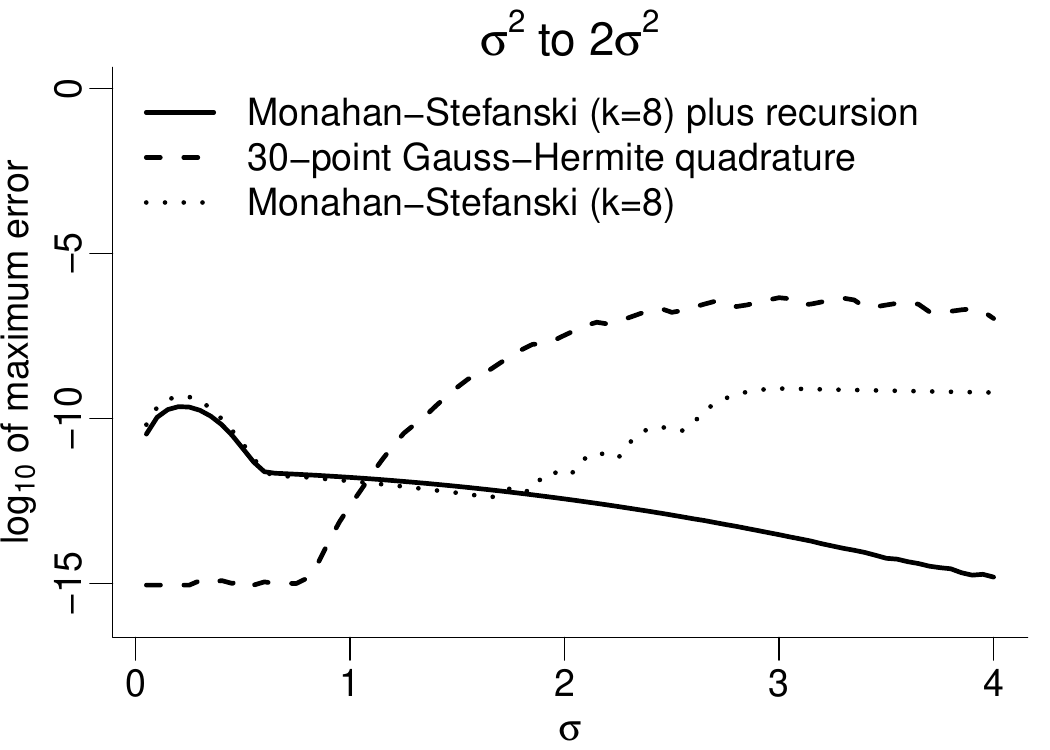}
\includegraphics[width=3.1in]{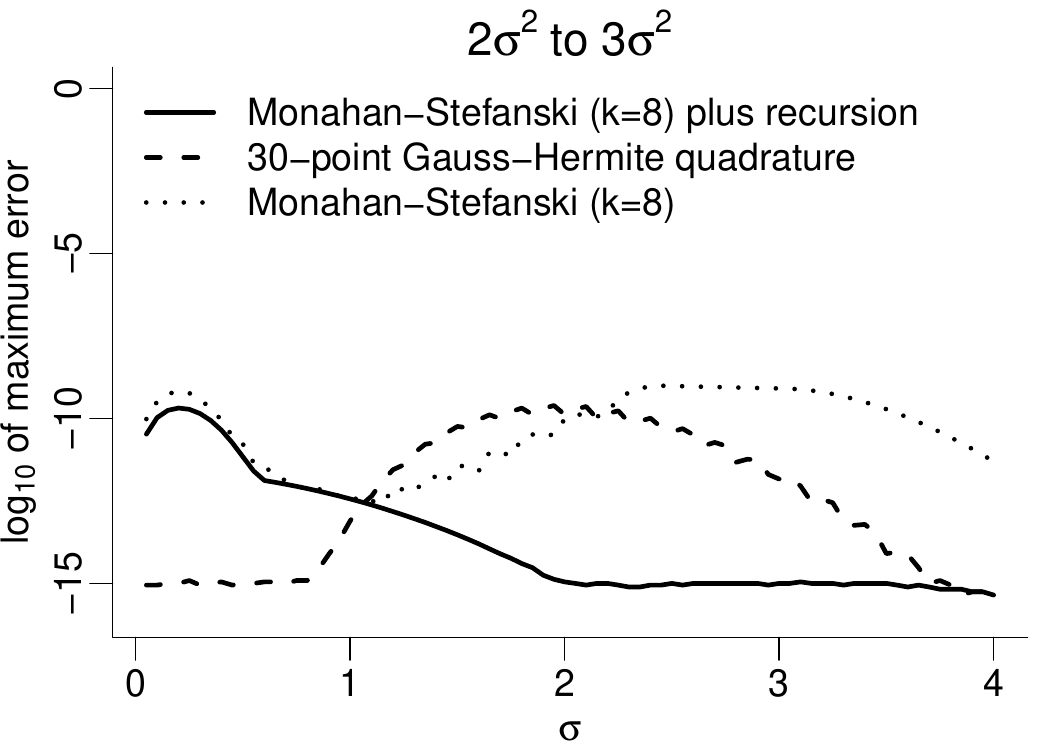}
\end{center}
\vspace{-0.2in}
\caption{\label{intgrlerrors} Maximum error relative to 1,000-point Gauss-Hermite quadrature for various integral approximations in the intervals $[\sigma^2,2\sigma^2]$ (left panel) and $[2\sigma^2,3\sigma^2]$ (right panel).  Machine accuracy is approximately $10^{-16}$, accounting for the floor in the two plots.} 
\end{figure}

For each of the competing methods and each value of $\sigma$ we computed the maximum ``error" relative to 1,000-point quadrature within each of the four intervals for $\mu$.  Figure~\ref{intgrlerrors} summarizes the results of the accuracy assessment for $[\sigma^2,2\sigma^2]$ and $[2\sigma^2,3\sigma^2]$.  Although 30-point Gauss-Hermite quadrature is the most accurate for small values of $\sigma$, our approach is the most accurate in the majority of cases.  Notably, our approach, which combines the recursion in (\ref{recursion}) with the approximation in (\ref{monahanapproxint}), clearly outperforms a direct application of (\ref{monahanapproxint}).

The 320,000 integral evaluations required for the accuracy assessment took 2.1 seconds for our approach compared to 2.1 seconds for the direct application of the Monahan-Stefanski method, 2.2 seconds for 30-point Gauss-Hermite quadrature, and 19.4 seconds for 1,000-point Gauss-Hermite quadrature.  Thus, the efficiency of our approach is comparable to that of the Monahan-Stefanski approach and slightly better than that of 30-point Gauss-Hermite quadrature.  Further, 1,000-point quadrature is considerably less efficient than the other three methods.  We conclude that our method offers the best tradeoff between accuracy and efficiency.

\vspace{-0.1in}
\subsection{Improving Mixing in the Presence of Many Random Effects} \label{sec:mixing}
\vskip-0.05in
\hspace{.9cm} A challenge associated with using MCMC to sample from a high-dimensional posterior density is poor mixing.  Due to the large number of unknown parameters, the only proposals that get accepted are those representing relatively small steps from the current state of the Markov chain.  Consequently, there is substantial autocorrelation in the Markov chain and it is necesssary to run the algorithm for an exceedingly long time to generate samples that represent approximately independent draws from the target posterior.

One way to improve mixing is to sample the parameters in blocks, but this may not always be enough.  We have found that when there are many random effects the latent random variables can dominate the likelihood and cause very few proposed $\boldsymbol{\beta}^*$ to be accepted.  To overcome this, we suggest that with each proposed $\boldsymbol{\beta}^*$ one simultaneously proposes random effects $\mathbf{U}^*$ that are consistent with the proposed fixed effects in order to increase the chance of acceptance.  An example illustrating this approach is given in Section~\ref{sec:epilepsydata}.  The $\boldsymbol{\beta}^*$ and $\mathbf{U}^*$ are proposed in such a manner that there is no net impact on the likelihood and the decision to accept or reject the proposed values is based entirely on the prior distributions.  A formal update of $\mathbf{U}$ is still required, but this strategy improves the acceptance rate for $\boldsymbol{\beta}$ and thereby facilitates faster mixing.

\vspace{-0.2in}
\section{Applications} \label{sec:applications}
\vskip-0.05in
\hspace{.9cm} In this section, we provide two examples of marginally interpretable GLMMs applied to the analysis of real data.  First, we use the technique described in Section~\ref{sec:logitnormal} to fit a model with a logit link to data from a teratological experiment on rats.  Next, we use the strategy introduced in Section~\ref{sec:mixing} to sample from the posterior of a marginally interpretable GLMM with a log link for data from a clinical trial of epileptics.  

\vspace{-0.2in}
\subsection{Rat Teratology} \label{sec:ratdata}
\vskip-0.05in
\hspace{.9cm} Our first example comes from a teratological experiment on rats conducted by \citet{weil1970}.  A group of 16 female rats was fed a diet containing a chemical agent during pregnancy and lactation, while another group of 16 female rats was fed a control diet.  Counts were made of the number of pups in each litter to survive four days from birth and to survive the 21-day lactation period.  Interest lies in the proportion of pups to survive 21 days among those alive after four days.  

For $i=1,2$ and $j=1,\dots,16$, we denote the number of pups in litter $j$ receiving treatment $i$ to survive four days by $m_{ij}$,  the number of pups to survive 21 days by $Y_{ij}$, and the proportion of pups to survive 21 days by $p_{ij}=Y_{ij}/m_{ij}$.  We include in our model a fixed effect for the treatment ($x_1=1$ for the treatment group, $x_2=-1$ for the control group) and random effects for litter (denoted $U_{ij}$).  Earlier analyses of these data established that there is more between-litter heterogeneity in the treatment group than in the control group \citep[see][]{lianghanfelt1994,heagertyzeger2000,wanglouis2004}. We therefore allow different random effects variances for the two treatment groups.  We assume $U_{ij} \overset{ind}{\sim} \mathrm{N}(0,\sigma_i^2)$ and $Y_{ij}|\boldsymbol{\beta},U_{ij} \overset{ind}{\sim} \mathrm{Binomial}\{m_{ij},\mathrm{E}(p_{ij}|\boldsymbol{\beta},U_{ij})\}$, define $h(\cdot)$ as the inverse logit function, and model the conditional mean as
\vskip-0.2in
\begin{equation*} \label{ratmodel}
\mathrm{E}(p_{ij}|\boldsymbol{\beta},U_{ij})= h(\beta_0 + \beta_1 x_i + U_{ij} + a_i),
\end{equation*} 
where $\boldsymbol{\beta}=(\beta_0,\beta_1)^T$ is the vector of fixed effects parameters and $a_i$ is the adjustment that ensures the model is marginally interpretable.  Since all litters in the same treatment group have the same covariate $x_i$ and the same random effects variance $\sigma_i^2$, they also have the same adjustment $a_i$.

We adopt a Bayesian approach and use MCMC to sample from the posterior distribution of the unknown parameters in our model.  Our prior distributions for $\beta_0$, $\beta_1$, $\log(\sigma_1^2)$, and $\log(\sigma_2^2)$ are $\mathrm{N}(0,25)$, $\mathrm{N}(0,10)$, $\mathrm{N}(-1/2,1)$, and $\mathrm{N}(-1/2,1)$, respectively.  To sample from our target posterior we iteratively update blocks of parameters using Metropolis steps.  We first update $\boldsymbol{\beta}=(\beta_0,\beta_1)^T$, then $\boldsymbol{\alpha}=(\sigma_1^2,\sigma_2^2)^T$, and finally $\mathbf{U}=(U_{1,1},\dots,U_{1,16},U_{2,1},\dots,U_{2,16})^T$.  When necessary, we compute the adjustment $a_i$ using the technique described in Section~\ref{sec:logitnormal}.  This MCMC algorithm was carried out both with the adjustment included in the model and without it.  Each chain was run for 1,010,000 steps, with the first 10,000 steps discarded as burn-in and every $100^{th}$ step thereafter retained for the final sample.  This resulted in 10,000 draws from the posterior distribution for each model.  Additional details regarding the algorithm are provided in the Supplementary Material.  

\begin{figure}[t]
\begin{center}
\includegraphics[width=3in]{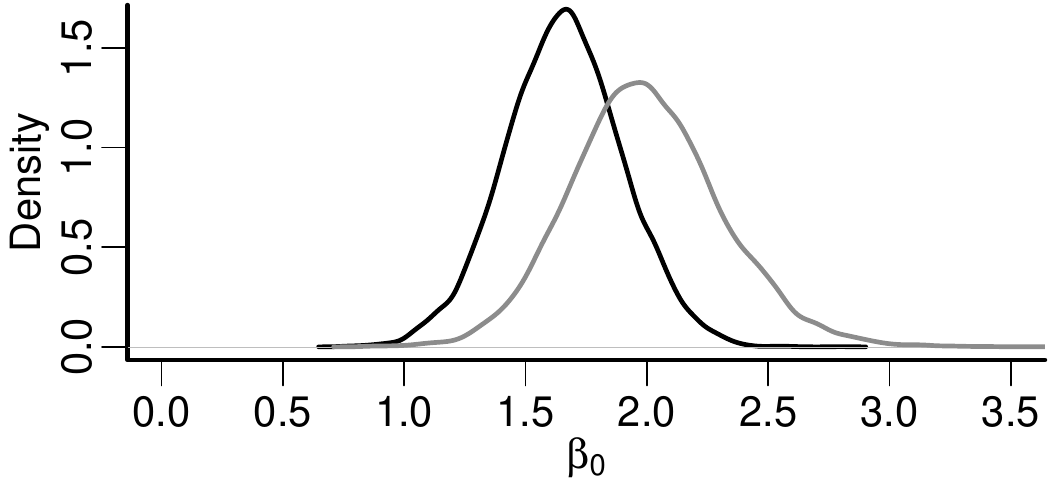}
\includegraphics[width=3in]{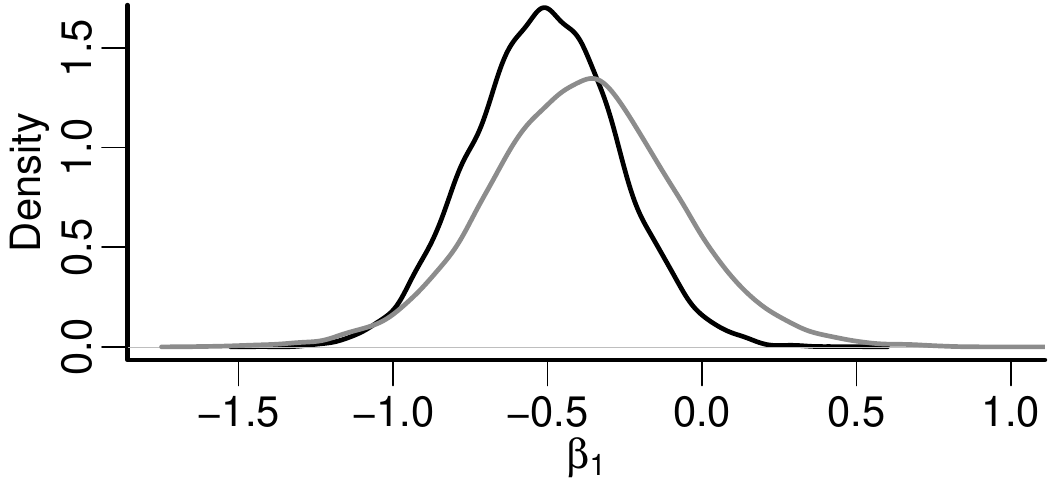}
\includegraphics[width=3in]{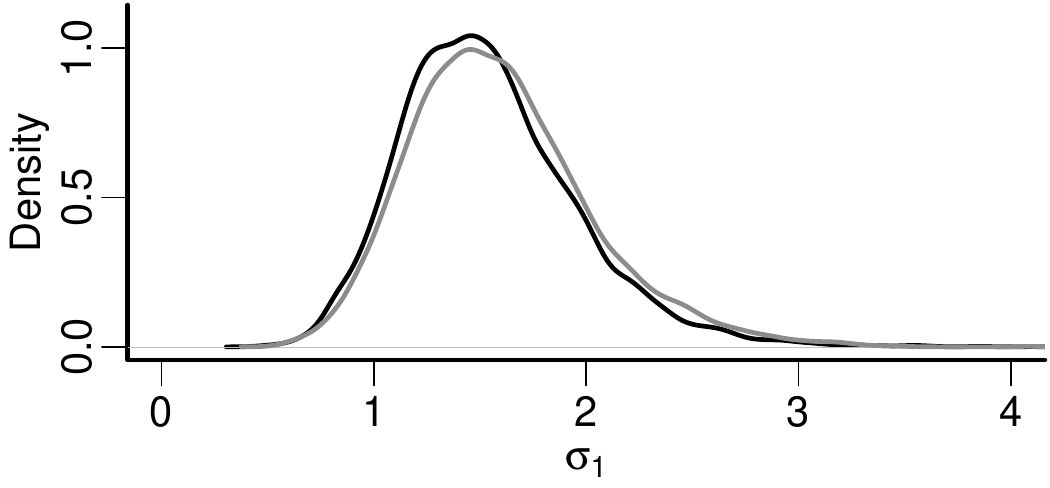}
\includegraphics[width=3in]{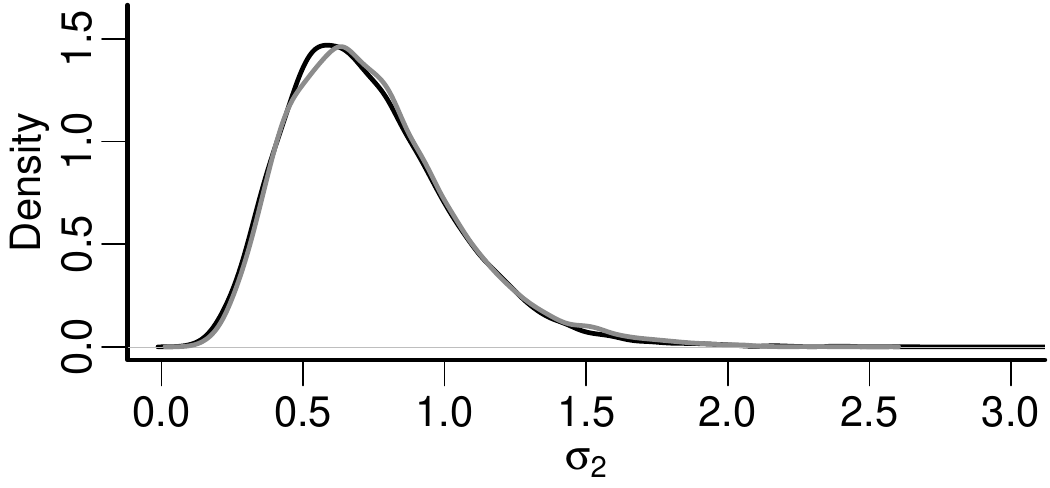}
\end{center}
\vspace{-0.31in}
\caption{\label{posteriordensities} Kernel density estimates of the posterior densities for the unknown parameters in the model for the rat teratology data.  Estimates obtained from the marginally interpretable model are in black while those obtained from the conventional GLMM are in gray.} 
\vspace{-0.06in}
\end{figure}

Table~\ref{ratests} provides posterior means and standard deviations for the parameters in both the marginally interpretable model and the conventional GLMM, which does not include the adjustment.  Figure~\ref{posteriordensities} displays kernel density estimates based on the posterior samples for the two models.  Notably, for $\beta_1$, which corresponds to the treatment effect, the tail area above zero is $0.016$ for the marginally interpretable model.  This is considerably less than the tail area of $0.101$ for the conventional GLMM.  Thus, many would draw different conclusions about the importance of the treatment effect using the two different models.  Indeed, Bayes factors for a test of no treatment effect ($H_0: \beta_1=0$), computed using the Savage-Dickey density ratio \citep[see][]{dickey1971, verdinelliwasserman1995}, favor the null hypothesis and come in at $1.27$ for the marginally interpretable model and $4.41$ for the conventional GLMM, thereby confirming this disparity.   

\begin{table}[t]
\small
\caption{\label{ratests} Posterior means of the unknown parameters in the model for the rat teratology data (with corresponding posterior standard deviations in parentheses)}
\vspace{-0.11in}
\begin{center}
\begin{tabular}{ccc}
Parameter &Marginally Interpretable GLMM &Conventional GLMM\\ \hline
$\beta_0$	&1.66 (0.24)		&1.99 (0.31) \\ 
$\beta_1$	&-0.51 (0.23)	&-0.39 (0.31) \\ 
$\sigma_1$	&1.54 (0.41)		&1.60 (0.43)	\\ 
$\sigma_2$	&0.73 (0.29)		&0.75 (0.30)
\end{tabular}
\end{center}
\vspace{-0.31in}
\end{table}

Since more rat pups survive than do not, the log-odds of survival are generally positive and therefore fall in a region where the inverse link function $h(\cdot)$ is concave.  Consequently, in the presence of random effects, the marginal mean is pulled down relative to the conditional mean.  This explains why the marginally interpretable intercept is less than $\beta_0$ in the conventional GLMM.  Further, since the random effects variance is greater in the treatment group than in the control group, the concavity of $h(\cdot)$ has a greater impact on the treatment group.  This contributes to $\beta_1$, the coefficient for the treatment effect, having a lesser value in the marginally interpretable model than in the conventional GLMM.  Note that the marginally interpretable $\beta_1$ is not attenuated toward zero relative to $\beta_1$ from the conventional GLMM, as one might expect, because the random intercept is not independent of $x_i$.

\begin{figure}[!h]
\begin{center}
\includegraphics[width=4.9in]{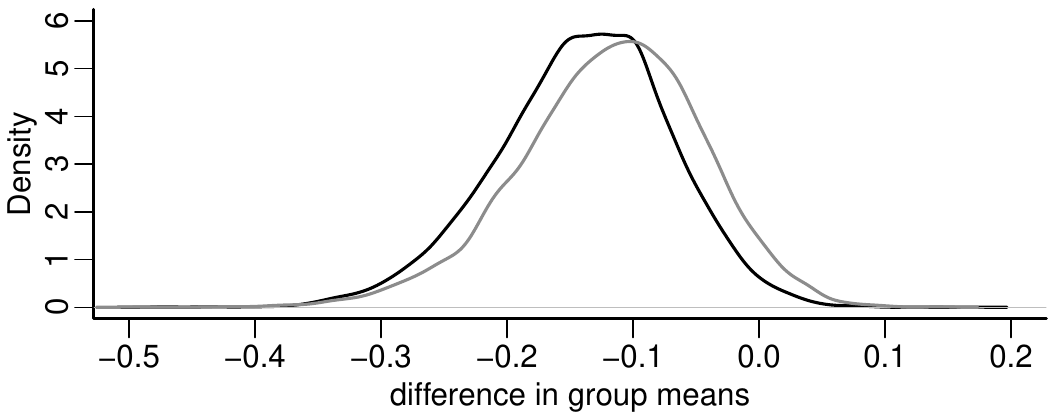}
\end{center}
\vspace{-0.3in}
\caption{\label{meanposteriors} Kernel density estimates of the posterior density of the difference in the expected 21-day survival rate between the two treatment groups based on the marginally interpretable model (black) and the conventional GLMM (gray).} 
\vspace{-0.05in}
\end{figure}

We compare the expected 21-day survival rates between the two treatment groups.  For the marginally interpretable model, the expected porportion of rat pups in the treatment group to survive 21 days among those alive after four days is $\mathrm{E}(p|\boldsymbol{\beta},\boldsymbol{\alpha},x=1)=h(\beta_0+\beta_1)$ and in the control group is $\mathrm{E}(p|\boldsymbol{\beta},\boldsymbol{\alpha},x=-1)=h(\beta_0-\beta_1)$.  For the conventional GLMM the same expectation is $\mathrm{E}(p|\boldsymbol{\beta},\boldsymbol{\alpha},x=1)=\int h(\beta_0+\beta_1+u)f_U(u)du$ for the treatment group and $\mathrm{E}(p|\boldsymbol{\beta},\boldsymbol{\alpha},x=-1)=\int h(\beta_0-\beta_1+u)f_U(u)du$ for the control group.  Note that the parameters $\boldsymbol{\alpha}$ enter this expression through the random effects distribution $f_U$.  Interest lies in whether or not the quantity $\mathrm{E}(p|\boldsymbol{\beta},\boldsymbol{\alpha},x=1)-\mathrm{E}(p|\boldsymbol{\beta},\boldsymbol{\alpha},x=-1)$ is nonzero.  Kernel density estimates of the posterior density for this quantity under the two models are shown in Figure~\ref{meanposteriors}.  The integral evaluation required for the conventional GLMM was accomplished using Monte Carlo integration.  Under both models, most of the posterior mass is below zero.  For the marginally interpretable model, the tail area above zero is 0.016, matching the tail area for $\beta_1$.  However, for the conventional GLMM, the tail area above zero is 0.041, which contrasts sharply with the tail area of 0.101 for $\beta_1$.

\vspace{-0.18in}
\subsection{Epileptic Seizures} \label{sec:epilepsydata}
\vskip -0.07 in
\hspace{.9cm} Our second example comes from a clinical trial of 59 epileptics conducted by \citet{leppiketal1987}.  Each subject received either a placebo or the drug progabide and then made four successive follow-up visits to the clinic during which they reported the number of partial seizures they had suffered in the two-week period immediately preceding the visit. We denote these reported counts by $Y_{ij}$, where $i=1,\dots,59$ indexes the subjects and $j=1,2,3,4$ indexes the visits.  \citet{thallvail1990} used GEE to fit a marginal model to these data.  They included as predictors the logarithm of one-fourth of the baseline count of partial seizures suffered by each patient in the eight-week period prior to treatment (denoted $\mbox{BASE}_i$), a treatment indicator ($1$ if progabide, $0$ if placebo, denoted $\mbox{TRT}_i$), the interaction between $\mbox{BASE}_i$ and $\mbox{TRT}_i$, the logarithm of the subject's age in years (denoted $\mbox{AGE}_i$), and a fourth-visit indicator ($1$ for the subject's fourth post-treatment visit, $0$ otherwise, denoted $\mbox{VISIT4}_j$).  \citet{breslowclayton1993} and \citet{gamerman1997} fit a GLMM with the same fixed effects and also two levels of random effects.  Their model has the form 
\vskip -0.4 in
\begin{equation} \begin{split} \label{epimodel}
\mathrm{E}(Y_{ij}|\boldsymbol{\beta},\gamma_i,\delta_{ij})= \exp\!\big{(}&\beta_0 + \beta_1 \times (\mbox{BASE}_i) + \beta_2 \times (\mbox{TRT}_i) + \beta_3 \times (\mbox{BASE}_i*\mbox{TRT}_i) + 
\\ &\beta_4 \times (\mbox{AGE}_i) + \beta_5 \times (\mbox{VISIT4}_j) + \gamma_i + \delta_{ij}\big{)},
\end{split} \end{equation} 
\vskip -0.1 in \noindent
where $\boldsymbol{\beta}=(\beta_0,\beta_1,\dots,\beta_5)^T$ is the vector of fixed effects parameters, the $\gamma_i \overset{ind}{\sim} \mathrm{N}(0,\sigma^2)$ are random subject effects, and the $\delta_{ij} \overset{ind}{\sim} \mathrm{N}(0,\tau^2)$ are random effects for visit within subject.  Further, conditional on the random effects $\gamma_i$ and $\delta_{ij}$, the reported seizure counts are assumed to be independent observations from a $\mathrm{Poisson}\{\mathrm{E}(Y_{ij}|\boldsymbol{\beta},\gamma_i,\delta_{ij})\}$.  \citet{breslowclayton1993} fit this model using PQL while \citet{gamerman1997} used MCMC to sample from a Bayesian version of this model.

We adopt a Bayesian approach and sample from a mixed model analogous to (\ref{epimodel}), but include an adjustment to ensure that the model is marginally interpretable.  In light of Section~\ref{sec:loglink}, the adjustment is simply $a_{ij}=-\sigma^2/2-\tau^2/2$ for all $i$ and $j$. We place $\mathrm{N}_6(\mathbf{0},100\mathbf{I}_6)$, $\mathrm{N}(-1,2)$, and $\mathrm{N}(-1,2)$ prior distributions on $\boldsymbol{\beta}$, $\log(\sigma^2)$, and $\log(\tau^2)$, respectively.  To sample from our target posterior using MCMC, the vector of parameters we must update is $\boldsymbol{\theta}=(\boldsymbol{\beta},\boldsymbol{\alpha},\boldsymbol{\gamma},\boldsymbol{\delta})^T$, where $\boldsymbol{\beta}$ is the vector of fixed effects parameters, $\boldsymbol{\alpha}=(\sigma^2,\tau^2)^T$ includes the parameters characterizing the random effects distribution, $\boldsymbol{\gamma}=(\gamma_1,\dots,\gamma_{59})^T$ includes the 59 latent variables associated with the subject random effect, and $\boldsymbol{\delta}=(\delta_{1,1},\dots,\delta_{59,4})^T$ includes the 236 latent variables associated with the visit random effect.  Due to the presence of 295 latent variables in this model, proposals for $\boldsymbol{\beta}$ are rarely  accepted when we use a basic MCMC algorithm that employs Metropolis steps to update the parameters in blocks.  

To address the problem with slow mixing, we simultaneously propose $\boldsymbol{\gamma}^*$ and $\boldsymbol{\delta}^*$ to be consistent with each proposed $\boldsymbol{\beta}^*$ as described is Section~\ref{sec:mixing}.  Specifically, for each $\boldsymbol{\beta}^*$ we also propose the following $\gamma_i^*$ and $\delta_{ij}^*$ for each $i=1,\dots,59$ and $j=1,2,3,4$: 
\vskip -0.2 in
\begin{equation} \label{gammashift}
\gamma_i^*=\gamma_i^{(t)} + \mathbf{x}_{i,1}^T(\boldsymbol{\beta}^{(t)}-\boldsymbol{\beta}^*),
\end{equation}
\begin{equation} \label{deltashift}
\delta_{i,4}^*=\delta_{i,4}^{(t)} + (\mathbf{x}_{i,4}^T-\mathbf{x}_{i,1}^T)(\boldsymbol{\beta}^{(t)}-\boldsymbol{\beta}^*), \ \mbox{and} \ \delta_{ij}^*=\delta_{ij}^{(t)} \ \mbox{for} \ j=1,2,3.  
\end{equation}
We then choose to accept or reject $\boldsymbol{\beta}^*$, $\boldsymbol{\gamma}^*$, and $\boldsymbol{\delta}^*$ collectively and set $\boldsymbol{\beta}^{(t+1)}$, $\boldsymbol{\gamma}'$, and $\boldsymbol{\delta}'$ accordingly.  The intermediate states $\boldsymbol{\gamma}'$ and $\boldsymbol{\delta}'$ are used in place of $\boldsymbol{\gamma}^{(t)}$ and $\boldsymbol{\delta}^{(t)}$ until $\boldsymbol{\gamma}$ and $\boldsymbol{\delta}$ are formally updated.  Proposing random effects to be consistent with the fixed effects in this manner increases the acceptance rate for $\boldsymbol{\beta}$ from $13.0\%$ to $51.2\%$.  Using this improved proposal scheme also decreases the integrated autocorrelation time for $\beta_3$, which corresponds to the interaction effect and has the highest such value among the six fixed effects parameters, from $580.0$ to $165.6$.  We carried out this MCMC algorithm both for the marginally interpretable model and the conventional GLMM.  We ran each chain for 2,100,000 steps, discarding the first 100,000 steps as burn-in and retaining every $200^{th}$ step thereafter to obtain a final sample of 10,000 draws from the posterior distribution for each model.  Additional details are provided in the Supplementary Material.  

Table~\ref{paramests} displays parameter estimates with corresponding measures of uncertainty for the marginally interpretable model and the conventional GLMM.  With the exception of the intercept $\beta_0$, the two sets of parameter estimates are virtually identical.  \citet{breslowclayton1993} noted that the slope parameters in this model have both a marginal and conditional interpretation while \citet{ritzspiegelman2004} stated that this will generally be the case for a model with a log link and a random intercept that is independent of the covariates in the model.  The intercept for the marginally interpretable model is greater than the intercept for the conventional GLMM due to the tendency of the convex inverse link function to pull the marginal mean up.

\begin{table}[t]
\small
\caption{\label{paramests} Posterior means of the unknown parameters in the model for the epilepsy data (with corresponding posterior standard deviations in parentheses)}
\vspace{-0.05in}
\begin{center}
\begin{tabular}{ccc}
Parameter &Marginally Interpretable GLMM &Conventional GLMM  \\ \hline
$\beta_0$	&-1.19 (1.23)	&-1.38 (1.23) \\ 
$\beta_1$	&0.88 (0.14)		&0.88 (0.14)	\\ 
$\beta_2$	&-0.95 (0.42)	&-0.96 (0.43) \\ 
$\beta_3$	&0.35 (0.22)		&0.35 (0.22)	\\ 
$\beta_4$	&0.48 (0.36)		&0.48 (0.36)	\\ 
$\beta_5$	&-0.10 (0.09)	&-0.10 (0.09) \\
$\sigma$	&0.50 (0.07)		&0.50 (0.07)	\\ 
$\tau$		&0.37 (0.04)		&0.37 (0.04)	
\end{tabular}
\end{center}
\vspace{-0.2in}
\end{table}

What separates the model with the adjustment from the conventional GLMM is its marginal interpretation.  Suppose we were interested in the average expected seizure count across all subjects in the population with a particular set of covariates.  For the marginally interpretable model, the quantity of interest is simply $\mathrm{E}(Y|\boldsymbol{\beta},\boldsymbol{\alpha}) = \exp(\mathbf{x}^T\boldsymbol{\beta})$.  Note that this quantity does not functionally depend on the parameters $\boldsymbol{\alpha}$ that characterize the random effects distribution.  For the conventional GLMM, the marginal mean is $\mathrm{E}(Y|\boldsymbol{\beta},\boldsymbol{\alpha}) = \int \exp(\mathbf{x}^T\boldsymbol{\beta} + \mathbf{d}^T\mathbf{u})f_\mathbf{U}(\mathbf{u})d\mathbf{u}$, which does depend on the random effects distribution.  Estimates of the fixed effects in a conventional GLMM are therefore more sensitive to the random effects than analogous estimates in a marginally interpretable model.  Fixed effects should be stable across different samples from the same population and perturbations of the random effects distribution should not impact them.  Thus, the marginally interpretable model is more generalizable to the entire population of interest than the conventional GLMM.  \citet{heagertykurland2001} made a similar point based on a simulation study investigating misspecification of the random effects distribution in a marginalized multilevel model.

A marginal model fit via GEE can also be used to make marginal inferences.  However, obtaining subject-specific predictions is considerably more difficult with such a model.  The marginally interpretable GLMM allows one to easily obtain both individual-level predictions and generalizable estimates of the marginal mean.  It provides a single, unified model that can be interpreted either marginally or conditionally depending on the goals of one's analysis.

\vspace{-0.23in}
\section{Conclusion and Discussion} \label{sec:discussion}
\vskip-0.07in
\hspace{.9cm} In this article we have defined a class of marginally interpretable GLMMs and described the form of the adjustment that appears in these models for several commonly-used link functions.  Unlike conventional GLMMs, which must be interpreted conditional on the random effects, these marginally interpretable GLMMs preserve the marginal mean even when the link function is nonlinear.  Consequently, model parameters can be given a \emph{population-averaged} interpretation.  In this sense, marginally interpretable GLMMs are comparable to marginal models fit via GEE, but unlike a purely marginal model, a marginally interpretable GLMM is a fully-specified model with a density for the data that can be used to make individual-level predictions in addition to marginal inferences.  We have also provided details regarding how to fit marginally interpretable GLMMs, including a fast and accurate algorithm for computing the logistic-normal integral.

Many of the examples we have provided relate to models with normal random effects, in part because the normal distribution is a common choice for random effects distributions.  However, a marginally interpretable GLMM does not require normal random effects and the methods described here apply to a wide array of random effects distributions.  One interesting class of random effects distributions consists of mixtures of normal distributions.  Mixed models that represent the random effects distribution as a mixture of normals \citep[see][]{magderzeger1996, caffoetal2007} allow considerable flexibility in the shape of the random effects distribution.
  
In this article we have focused on deriving the adjustment in a marginally interpretable GLMM by relating the conditional mean to the marginal mean.  Acknowledging that the marginal mean may not always be of interest, we can define marginally interpretable models in other mixed model settings.  For example, in mixed-effect quantile regression models \citep[see][]{koenker2004,geracibottai2014} or when modeling extremes \citep[see][]{coles2001,stephensontawn2004} we can consider a definition for a marginally interpretable model based on relating the conditional quantiles to the marginal quantiles.  Further research is needed to understand the form of the adjustments that arise in these settings.

Although we discuss fitting marginally interpretable GLMMs using Bayesian techniques, these models are also compatible with frequentist techniques.  Regardless of the method used, the key is to include the adjustment at the appropriate step in the algorithm.  Further, for models with a logit link and normal random effects, any model-fitting technique, be it Bayesian or frequentist, could benefit from our more accurate approach to evaluating the logistic-normal integral.

Another area where the marginally interpretable GLMM can improve inference is in hypothesis testing.  When comparing two group means in a GLMM with a nonlinear link function, testing whether there is a difference between the two groups is not necessarily the same as testing whether the group means differ because the different groups could require different adjustments to preserve the marginal means.  In other words, failure to account for the impact of the nonlinear link could lead one to test the wrong hypotheses.  The importance of the adjustment was demonstrated with the rat teratology data in Section~\ref{sec:ratdata}.  Using the conventional GLMM, testing for a nonzero treatment effect yielded a different result than testing for a difference in the expected survival rates between the two treatment groups. Using the marginally interpretable GLMM avoids such inconsistencies because it makes the appropriate adjustments in the presence of random effects.

\vspace{-0.23in}
\section*{Acknowledgments}
\vskip -0.07 in
Craigmile is supported in part by the US National Science Foundation under grants DMS-1407604 and SES-1424481, while MacEachern is supported in part by grant DMS-1613110.

\singlespacing
\bibliographystyle{apalike}
\bibliography{referencelist1}

\end{document}